\def\year{2021}\relax
\documentclass[letterpaper]{article} % DO NOT CHANGE THIS
\usepackage{aaai21}  % DO NOT CHANGE THIS
\usepackage{times}  % DO NOT CHANGE THIS
\usepackage{helvet} % DO NOT CHANGE THIS
\usepackage{courier}  % DO NOT CHANGE THIS
\usepackage[hyphens]{url}  % DO NOT CHANGE THIS
\usepackage{graphicx} % DO NOT CHANGE THIS
\usepackage{csquotes}
\usepackage{rotating}
\usepackage{natbib}  % DO NOT CHANGE THIS AND DO NOT ADD ANY OPTIONS TO IT
\usepackage{caption} % DO NOT CHANGE THIS AND DO NOT ADD ANY OPTIONS TO IT
\usepackage{array}
\usepackage[T1]{fontenc}
\title{Observational Study of Working from Home during the COVID-19 Pandemic Using Social Media Data}
\author{
    Ziyu Xiong, Pin Li, Hanjia Lyu, Jiebo Luo
    \\
}
\affiliations{
    University of Rochester\\
    \{zxiong7, pin.li, hlyu5\}@ur.rochester.edu,
    jluo@cs.rochester.edu
}
\begin{document}

\maketitle

\begin{abstract}
\textbf{Background:} Since March 2020, companies nationwide have started work from home (WFH) due to the rapid increase of confirmed COVID-19 cases in an attempt to help prevent the coronavirus from spreading and rescue the economy from the pandemic. Many organizations have conducted surveys to understand people's opinions towards WFH. However, the findings are limited due to a small sample size and the dynamic topics over time.

\textbf{Objective:} The study aims to understand the U.S. public opinions on working from home during the COVID-19 pandemic.

\textbf{Methods:} We conduct a large-scale social media study using Twitter data to portrait different groups who have positive/negative opinions about WFH. We perform an ordinary least squares regression to investigate the relationship between the sentiment about WFH and user characteristics including gender, age, ethnicity, median household income, and population density. To better understand public opinion, we use latent Dirichlet allocation to extract topics and discover how tweet contents relate to people's attitude.

\textbf{Results:} After performing the ordinary least squares regression using a large-scale dataset (N = 28,579) of publicly available Twitter posts concerning working from home ranging from April 10, 2020 to April 22, 2020, we confirm that sentiment of working from home varies across user characteristics. In particular, women tend to be more positive about working from home (p < .001). People in their 40s are more positive towards WFH than other age groups (p < .001). People from high-income areas are more likely to have positive opinions about working from home (p < .001). These nuanced differences are supported by a more fine-grained topic analysis. At a higher level, we find that the most negative sentiment about WFH roughly corresponds to the discussion of government policy. However, people express more positive sentiment when talking about topics on “remote work/study” and “encouragement”. Furthermore, topic distributions vary across different user groups. Women pay more attention to family activities than men (p < .05). Older people talk more about work and express more positive sentiment on WFH.

\textbf{Conclusions:} This paper presents a large-scale social media-based study to understand the U.S. public opinions on working from home during the COVID-19 pandemic. We hope this study can lend itself to making policies both at national and institution/company levels to improve the overall population's experience of working from home.

\end{abstract}

\section{Introduction}
\subsection{Background}
COVID-19, also known as the coronavirus, first reported in China and then spread to the whole world, has caused 22.3 million confirmed cases and more than 373 thousand deaths in the U.S. by January 11, 2021\footnote{https://covid.cdc.gov/covid-data-tracker [Accessed March 25, 2021]}. To help prevent the virus from spreading and also salvage the economy, companies and schools nationwide have started work/study from home. According to a Gartner survey of 880 global HR executives on March 17, 2020, almost 88\% organizations have encouraged or required employees to work from home\footnote{https://www.gartner.com/en/newsroom/press-releases/2020-03-19-gartner-hr-survey-reveals-88--of-organizations-have-e [Accessed March 25, 2021]}. \citet{barrero2020working} have found that working from home might stick even after the pandemic ends. Concerns may arise when it comes to productivity~\cite{feng2020covid}, willingness~\cite{palumbo2020let}, and future trends~\cite{barrero2020working, chung2020working} regarding work/study from home. 

\subsection{Prior Work}
Working from home has been a controversial issue which merits a closer look. An investigation shows that WFH might incur side-effects such as a negative impact on work-life balance~\cite{palumbo2020let}. This would lead to negative opinions towards WFH when people tweet about it. Other research deep dive into specific categories. A survey of Lithuania's employees shows that female employees appreciate more than male employees, because the female employees can enjoy a healthier lifestyle while male employees worry about career constraints~\cite{raivsiene2020working}. However, another survey conducted in the U.S. shows “a gender gap in perceived work productivity”: Before WFH, female and male employees report the same level of self-rated work productivity. After shifting to WFH, male employees perform with better productivity than the female employees~\cite{feng2020covid}. As for age, people in their 40s have more negative opinions on WFH because of their unfamiliarity with teleworking. People aged 30-39 have the most positive opinions because they can enjoy time with family and they are already used to new technologies for teleworking~\cite{raivsiene2020working}. Previous studies~\cite{barrero2020working, raivsiene2020working, bick2020work} also show that opinions concerning working from home vary across different socioeconomic groups. A similar social media study of public sentiment on working from home has been conducted in the U.K.~\cite{carrollcovid}. Results show that in the U.K., more than 70\% tweets concerning working from home have positive sentiment and the main topics include traffic, drink, and e-Commerce. 

Similar approaches have been implemented by researchers mining Twitter posts with natural language processing on their attitudes on face masks~\cite{yeung2020face}, using the Valence Aware Dictionary and Sentiment Reasoner (VADER) model~\cite{hutto2014vader} to perform sentiment analysis. Moreover, Twitter data have been used to study many different aspects of COVID-19, such as mining overall public perception towards COVID-19~\cite{boon2020public}, college students' attitudes on the pandemic~\cite{duong2020ivory}, people's attitude on potential COVID-19 vaccines~\cite{Lyu2020.12.12.20248070, wu2021characterizing}, pregnant women sentiment analysis during quarantine~\cite{talbot2021feeling}, as well as monitoring depression trend on Twitter during the COVID-19 pandemic~\cite{zhang2021monitoring}. They all use VADER for sentiment analysis and most of them include time series analysis. In addition, we follow the practice of using Latent Dirichlet Allocation (LDA)~\cite{blei2003latent}, to identify topics among large text corpus. The M3-inference Model is used by \citet{yeung2020face} to portrait different demographic groups.

\subsection{Goal of the Study}
In this study, we intend to understand public opinions on work from home using large-scale social media data. Twitter has been a popular social media platform for people, especially in the U.S., to express their feelings about what is happening around them. In contrast, the Boston Consulting Group used survey data to study employees' opinions regarding COVID-19 work from home\footnote{https://www.bcg.com/publications/2020/valuable-productivity-gains-covid-19 [Accessed March 25, 2021]}. However, social media data allows an opportunity for conducting a more timely study of many population-level issues on a larger scale~\cite{jin2010wisdom}. We acquire the data with an authorized Twitter developer account using Tweepy. This ensures reliability by acquiring first-hand and sufficient data when conducting the research.

In this paper, we also infer user demographic information using Twitter user information. This is of importance since we can deep dive into the characteristics of those who are more pro-WFH. For example, when we look into gender, we understand that historically mothers have been mostly responsible for caring for children\footnote{https://www.ucpress.edu/book/9780520246461/from-marriage-to-the-market [Accessed March 27, 2021]}. Therefore, we need the gender information to check whether or not there is any difference in sentiment towards working from home between women and men, as working from home would allow female employees to allocate more time accompanying their children. 

Our goal is to understand the U.S. public opinions on working from home during the COVID-19 pandemic. In particular, we focus on the following research questions:

\begin{itemize}
    \item \textbf{RQ1:} Who is more likely to tweet about working from home?
    \item \textbf{RQ2:} How does the sentiment of working from home vary across user demographics?
    \item \textbf{RQ3:} When discussing working from home, what do Twitter users mainly talk about? How does the content correlate with the sentiment of working from home?
\end{itemize}

To summarize, in a large-scale dataset of publicly available Twitter posts concerning working from home ranging from April 10, 2020 to April 22, 2020, we find that women and older people are more likely to tweet about working from home. After performing the ordinary least squares regression, we confirm that sentiment of working from home varies across user characteristics. In particular, women tend to be more positive about working from home. People in their 40s are more positive towards WFH than other age groups. People from high-income areas are more likely to have positive opinions about working from home.

These nuanced differences are supported by a more fine-grained topic analysis. At a higher level, we find that the most negative sentiment about WFH roughly corresponds to the discussion of government policy. However, people express more positive sentiment when talking about topics on “remote work/study” and “encouragement”. Furthermore, topic distributions vary across different user groups. 

\section{Methods}
In this section, we summarize the data collection process and the methods we apply in the analyses. To address \textbf{RQ1} and \textbf{RQ2}, we discuss how we infer user characteristics and the sentiment in \textbf{Feature Inference}. To investigate \textbf{RQ3}, we describe how we extract the topics of tweets in \textbf{Topic Modeling}.

\subsection{Data Collection}
We collect related English tweets through Tweepy stream API using keywords and hashtags filtering. The filter keywords and hashtags are “WFH”, “workfromhome”, “work from home”, “\#wfh”, “\#workingfromhome”.  553,166 unique tweets with 23 attributes posted by 405,455 unique Twitter users ranging from April 5, 2020 to April 26, 2020 are collected. We attempt to infer the gender, age, ethnicity of the Twitter users, extract the population density of the location, and estimate the sentiment of the tweets. There are 405,455 unique users in our dataset, 313,815 of them (77.3\%) only tweeted once. After removing duplicates and the users with incomplete features, 28,579 unique Twitter users with all features are included in the dataset.

\subsection{Feature Inference}
\subsubsection{Sentiment.}
A normalized, weighted composite score is calculated for each tweet using VADER (Valence Aware Dictionary for sEntiment Reasoning)~\cite{hutto2014vader} to measure the sentiment. The score ranges from -1 (most negative) to +1 (most positive). As validation, we randomly select 194 users’ tweets within one month. By manually labeling the sentiment and comparing the sentiment scores with the VADER scores, we find that the accuracy is 76\%, suggesting that the automatic NLP (Natural Language Processing) methods we employ provide adequate estimates of the sentiment of the tweets. Table~\ref{tab:des_sentiment} shows the descriptive statistics of the sentiment score.

\begin{table}[htbp]
    \centering
    \scriptsize
    \caption{Descriptive statistics of sentiment score.}
    \begin{tabular}{c c c c c c c}
    \hline
    Mean & Std & Min & 25\% & 50\% & 75\% & Max\\
    \hline
    0.242 & 0.448 & -0.967 & 0.000 & 0.318 & 0.617 & 0.984\\
 \hline
    \end{tabular}
    \label{tab:des_sentiment}
\end{table}

\subsubsection{Age and Gender.} We apply the M3-inference model~\cite{wang2019demographic} to infer the gender and age of each Twitter user using profile name, user name (screen name), and profile description. Age is binned into four groups: <=18, 19-29, 30-39, >=40. The gender distribution of Twitter users is biased on men around 71.8\%~\cite{burger2011discriminating}. A similar pattern is also observed in our dataset, where 57.9\% are men, and 42.1\% are women. With respect to age, 37.08\% of the users in our dataset are older than 40 years old, 37.6\% are between 30 to 39 years old, 16.5\% are between 19 to 29 years old, and the rest are younger than 19 years old. According to a report of the Pew Research Center\footnote{https://www.pewresearch.org/internet/2019/04/24/sizing-up-twitter-users/ [Accessed June 4, 2021]}, Twitter users are younger than the average U.S. adult. 21\% are those aged 18-29, 33\% aged 30-49, 26\% aged 50-64, and 20\% aged 65 and older. The percentages of adults of the Twitter population are 29\%, 44\%, 19\% and 8\%, respectively. The pattern in our dataset is more similar to the U.S. adult distribution.

\subsubsection{Ethnicity.} To estimate the ethnicity of the Twitter users, we apply the Ethnicolr API which makes inference based on the last name and first name or just the last name of the Twitter user~\cite{sood2018predicting}. In our work, we remove emoji icons, hyphens, unrelated contents and special characters to extract the last names and apply “census\_ln” to infer the ethnicity which contains \textit{White}, \textit{Black or African American}, \textit{Asian/Pacific Islander}, \textit{American Indian/Alaskan Native}, and \textit{Hispanic}. In our dataset, \textit{White} is predominant over other categories with 83.4\%, while according to the U.S. Census Bureau~\cite{census2019}, \textit{White} constitutes 60.1\% of the U.S. population; 7.3\% are \textit{Asian/Pacific Islander}, while \textit{Asian/Pacific Islander} constitutes 6.1\% in the U.S.; 6.5\% are \textit{Hispanic}, while 18.5\% of the U.S. population are \textit{Hispanic}; 2.5\% are \textit{Black or African American}, while 13.4\% of the U.S. population are \textit{Black or African American}; \textit{American Indian/Alaska Native} only constitutes 0.26\%. According to the report of the Pew Research Center, the percentages in race/ethnicity are almost the same between U.S. adults and Twitter adult users. Interestingly, the percentages of \textit{White} and \textit{Asian/Pacific Islander} are much higher than those in the general population, which could be related to the labor force distributions of these two groups. In 2018, 54\% of employed \textit{Asian} and 41\% of employed \textit{White}, compared with 31\% of employed \textit{Black or African American} and 22\% of employed \textit{Hispanic} worked in management, professional, and related occupations\footnote{https://www.bls.gov/opub/reports/race-and-ethnicity/2018/home.htm [Accessed March 25, 2021]} that can be most likely done at home~\cite{dingel2020many}. Therefore, it is not surprising that there are more \textit{White} and \textit{Asian/Pacific Islander} in our dataset due to the disparities in the occupations.

\subsubsection{Population density.} USzipcode SearchEngine is applied to extract the population density of each user's location that is self-reported by the Twitter user in the profile information. The population density is categorized into urban (greater than 3,000), suburban (1,000-3,000) and rural (lower than 1,000). In the end, 67.4\% are urban, 14.6\% are suburban and the rest are rural. The majority of the users of our dataset are from urban areas, which is consistent with the fact that 83\% of the U.S. population lived in urban areas\footnote{http://css.umich.edu/factsheets/us-cities-factsheet [Accessed March 27, 2021]}, however, there are proportionally fewer urban users in our dataset than in the U.S. population.

\subsubsection{Income.} To understand the relationship between people's attitude towards working from home and the gap between rich and poor at users' locations, we retrieve regional median income from 2019 American Community Survey (ACS). Census Application Programming Interface (API) tools are used to extract median income with an input of city-level user location. The descriptive statistics are shown in Table~\ref{tab:des_income}.

\begin{table}[htbp]
    \centering
    \scriptsize
    \caption{Descriptive Statistics of the regional median income.}
    \begin{tabular}{c c c c c c c}
    \hline
    Mean & Std & Min & 25\% & 50\% & 75\% & Max\\
    \hline
    33,538 & 10,298 & 3,951 & 28,072 & 31,613 & 36,336 & 121,797\\
 \hline
    \end{tabular}
    \label{tab:des_income}
\end{table}

\subsection{Topic Modeling}
We use LDA~\cite{blei2003latent} to extract topics from the tweets. In our study, we use the stop words package from NLTK library, extended with topic related words (e.g., “work”, “home”). To extract the most relevant topics, we only collect nouns, verbs, adjectives and adverbs lemmas. We use the spaCy package to go through all the words of the tweets, and only include the words whose postag is “NOUN”, “ADJ”, “VERB” or “ADV”. We tune the hyperparameters by nested looping topic numbers, $\alpha$ and $\beta$. In the end, we choose num\_topics=9, $\alpha = 0.91$, $\beta = 0.31$, with a coherence score $C_{v}=0.379$.

\section{Results}
\subsection{Sentiment Analysis}
In the previous section, we find that when referring to working from home, Twitter users are slightly positive. In this section, we attempt to investigate the relationship between user characteristics and the sentiment of discussions about working from home. We perform an ordinary least squares regression on the dataset n=28,579. Descriptive statistics and bi-variate correlations are shown in Table~\ref{tab:char_desc}. Table~\ref{tab:regression_outputs} summarizes the result of the ordinary least squares regression.

\begin{table*}[htbp!]
\scriptsize
\centering
\caption{Descriptive statistics and the bi-variate correlations. Income is normalized by MinMaxScaler.}
% \tabcolsep=0.03cm
% \scriptsize
\setlength{\tabcolsep}{2pt} % Default value: 6pt
\begin{tabular}{l r r l l l l l l l l l l l l} 
\hline
\textbf{Variables} & \multicolumn{1}{c}{\textbf{Mean}} & \multicolumn{1}{c}{\textbf{SD}} & \multicolumn{1}{c}{\textbf{1}} &  \multicolumn{1}{c}{\textbf{2}} & \multicolumn{1}{c}{\textbf{3}} &\multicolumn{1}{c}{\textbf{4}} & \multicolumn{1}{c}{\textbf{5}} &\multicolumn{1}{c}{\textbf{6} }&\multicolumn{1}{c}{\textbf{7}} &\multicolumn{1}{c}{\textbf{8}} & \multicolumn{1}{c}{\textbf{9}} &\multicolumn{1}{c}{\textbf{10}} \\
\hline
%\multicolumn{7}{|c|}{VADER Compound Score} \\
%\hline
%   mean & std & min & 25\% & 50\% & 75\% & max\\
%   \hline
%   0.119 & 0.504 & -0.984 & -0.296 & 0.000 & 0.540 & 0.994 \\
%   \hline
1 Gender (0 = female, 1 = male) & 0.59 & 0.49\\

2 Age $<=18$ (0 = No, 1 = Yes) & 0.09 & 0.28 & -$.04^{**}$\\

3 Age 19-29 (0 = No, 1 = Yes) & 0.16 & 0.37 & -$.21^{**}$ & -$.14^{**}$\\

4 Age 30-39 (0 = No, 1 = Yes) & 0.38 & 0.48 & $\:.01$ & -$.24^{**}$ & -$.34^{**}$\\

5 \textit{Black or African American} (0 = No, 1 = Yes) & 0.03 & 0.16 & $\:.01^{*}$ & $\:.02^{**}$ & -$.00$ & -$.03^{**}$ \\

6 \textit{Asian/Pacific Islander} (0 = No, 1 = Yes) & 0.07 & 0.26 & $\:.01$ & $\:.07^{**}$ & -$.02^{**}$ & -$.01$ & -$.05^{**}$\\

8 \textit{American Indian/Alaska Native.} (0 = No, 1 = Yes) & 0.01 & 0.05 & $\:.02^{**}$ & $\:.03^{**}$ & -$.01$ & -$.00$ & -$.01$ & -$.01^{*}$\\

7 \textit{Hispanic} (0 = No, 1 = Yes) & 0.07 & 0.25 & -$.00$ & $\:.02^{**}$ & $\:.04^{**}$ & $\:.03^{**}$ & -$.04^{**}$ & $\:.00$ & -$.02^{**}$ \\

9 Income & 0.25 & 0.09 & $\:.04^{**}$ & -$.03^{**}$ & -$.07^{**}$ & -$.01$ & -$.01$ & $\:.04^{**}$ & $\:.00$ & -$.02^{**}$ & \\

10 Urban (0 = No, 1 = Yes) & 0.67 & 0.47 & -$.02^{**}$ & $\:.02^{**}$ & $\:.03^{**}$ & $\:.02^{**}$ & -$.01$ & $\:.05^{**}$ & -$.00$ & $\:.03^{**}$ & $\:.05^{**}$\\

11 Suburban (0 = No, 1 = Yes) & 0.15 & 0.35 & $\:.02^{**}$ & -$.02^{**}$ & -$.02^{**}$ & $\:.01^{*}$ & $\:.01^{*}$ & -$.02^{**}$ & $\:.00$ & $\:.00$ & $\:.09^{**}$ & -$.60$\\
\hline
\end{tabular}\\
    {\raggedright Note. * $p<.05$. ** $p<.01$. \par}    \label{tab:char_desc}
\end{table*}

\begin{table*}[]
    \centering
    \caption{Ordinary least squares regression outputs for public opinion on Working from Home against demographics and other variables of interest.}
    \setlength{\tabcolsep}{3pt} % Default value: 6pt
    \renewcommand{\arraystretch}{1.5} % Default value: 1
    \begin{tabular}{r l l l}
    \hline
     & \multicolumn{3}{l}{\textbf{Sentiment score}} \\
     \cline{2-4}
     
     %\hline
     Predictor & \textit{B} & \textit{SE} & 95\% CI \\
\hline
Intercept & $0.252^{***}$ &0.011 & (0.231, 0.273) \\
     Gender (0=Female, 1=Male) & -$0.021^{***}$ & 0.006 & (-0.032, -0.010) \\
Age $<=18$ (0=No, 1=Yes)& -$0.084^{***}$ &0.010& (-0.103, -0.064)\\
           Age 19-29 (0=No, 1=Yes) & -$0.076^{***}$  & 0.008 & (-0.092, -0.060)\\
           Age 30-39 (0=No, 1=Yes) & -$0.022^{***}$ & 0.005 & (-0.034, -0.010) \\
           \textit{Black or African American} (0=No, 1=Yes) & 0.023 & 0.017 & (-0.011, 0.066)\\
           \textit{Asian/Pacific Islander} (0=No, 1=Yes) & 0.003 & 0.010 & (-0.020, 0.020) \\
           \textit{American Indian/Alaska Native} (0=No, 1=Yes) & -0.006 & 0.011 & (-0.027, 0.016)\\
           \textit{Hispanic} (0=No, 1=Yes) & -0.006 & 0.011 & (-0.027, 0.016)\\
           Income & $0.143^{***}$ & 0.031 & (0.082, 0.203) \\
           Urban (0=No, 1=Yes) & -0.007 & 0.007 & (-0.021, 0.007) \\
           Suburban & -0.004 & 0.009 & (-0.023, 0.014) \\

          \hline
          F-statistic & \multicolumn{3}{c}{$15.25^{***}$}\\
        $R^{2}$ & \multicolumn{3}{c}{0.006}\\
        Adjusted $R^{2}$ & \multicolumn{3}{c}{0.005}\\
        Sample size & \multicolumn{3}{c}{28,579}\\          
          \hline
          {\raggedright Note. * $p<.05$. ** $p<.01$. *** $p<.001$.\par}
    \end{tabular}
    \label{tab:regression_outputs}
\end{table*}

\subsubsection{Women tend to be more positive about working from home.} Men are significantly more negative about working from home than women (p < .001). This is consistent with the remote work survey report by Fast Company\footnote{https://www.fastcompany.com/90477102/4-ways-remote-work-is-better-for-women [Accessed March 24, 2021]}. A more positive sentiment observed in women could be due to the change in working styles and fewer work hours compared to men~\cite{collins2020covid}. Previous survey indicates that women favor WFH from a healthier lifestyle perspective~\cite{raivsiene2020working}. 

\subsubsection{People in their 40s are more positive towards WFH than other age groups.} Age is another perspective. The regression results show that as ages increase, people are significantly more pro-WFH (p < .001).  This is consistent with the survey result conducted by Hannah Watkins~\footnote{https://hubblehq.com/blog/future-of-work-different-age-groups [Accessed March 25, 2021]} that Gen Z (people at the point of this report aged from 8-23) are more pro-office than Millennials (aged 24-39). While assumptions exist when it comes to the older employees, who might be unfamiliar with electronic devices and thus become more pro-office. However, according to an article in Financial Times\footnote{https://www.ft.com/content/46ac277c-da5f-4b99-838d-b0ee228d8c57 [Accessed June 9, 2021]}, the case is just the opposite. People aged 40s and older are less likely to be re-employed. Thus, they would like to keep their current jobs while avoiding the risk of getting exposed to COVID-19, especially since this group is most vulnerable to COVID-19. Further details about the topics will be discussed in the following section. It also shows the same pattern as the survey conducted in Lithuania~\cite{raivsiene2020working}.

\subsubsection{People from higher-income areas are more likely to have pro-WFH opinions than the people from lower-income areas.} Income is significantly correlated with the sentiment about working from home (p < .001). This aligns with our findings that people from urban areas would be more pro-WFH, since the regional median income would be higher in big cities\footnote{https://www.statista.com/statistics/205609/median-household-income-in-the-top-20-most-populated-cities-in-the-us/ [Accessed March 25, 2021]}. This is also in line with the findings of \citet{barrero2020working} that high-income workers, especially, enjoy the perks of working from home. 

\subsection{Topic Analysis}
Next, we attempt to capture what Twitter users mainly talk about when they refer to working from home. More specifically, we investigate how the contents correlate with the sentiment of working from home. Table~\ref{tab:topic} shows the 9 topics extracted by the LDA model. We assign each topic a title on the basis of the top 10 keywords.

\begin{table*}[htbp!]
    \centering
    \caption{Titles and the top 10 keywords of the topics extracted by LDA.}
    \begin{tabular}{l l l}
    \hline
    Topic & Topic Title & Topic Keywords \\
    \hline
    1     & Family activities & dog, try, today, wife, day, last, school, virtual, watch, look \\ 
    \hline
    2     & Remote work/study & remote, new, time, covid, learn, great, help, many, support, join\\ 
    \hline
    3     & Quarantine & pandemic, stay, safe, day, go, today, let, see, also, think\\ 
    \hline
    4     & Dressing & dress, get, enough, adult, day, wear, time, zoom, right, thank\\ 
    \hline
    5     & Government and policy & money, less, job, option, able, new, remotely, safely, force, take\\ 
    \hline
    6     & COVID side influence & people, still, do, job, good, say, first, place, probably, go\\ 
    \hline
    7     & Encouragement & get, lot, world, honestly, fall, reveal, love, week, look\\ 
    \hline
    8     & Back to office & go, back, know, office, time, feel, quarantine, hour, tip, covid\\ 
    \hline
    9     & Leasing  & office, well, year, instead, couple, permanently, think, lease, renew, similarly\\ 
    \hline
    % 10    & General & get, may, lot, honestly, low, world, reveal, day, try, good\\ 
    % \hline
    \end{tabular}
    \label{tab:topic}
\end{table*}

\begin{figure}[htbp]
    \centering
    \includegraphics[width = \linewidth]{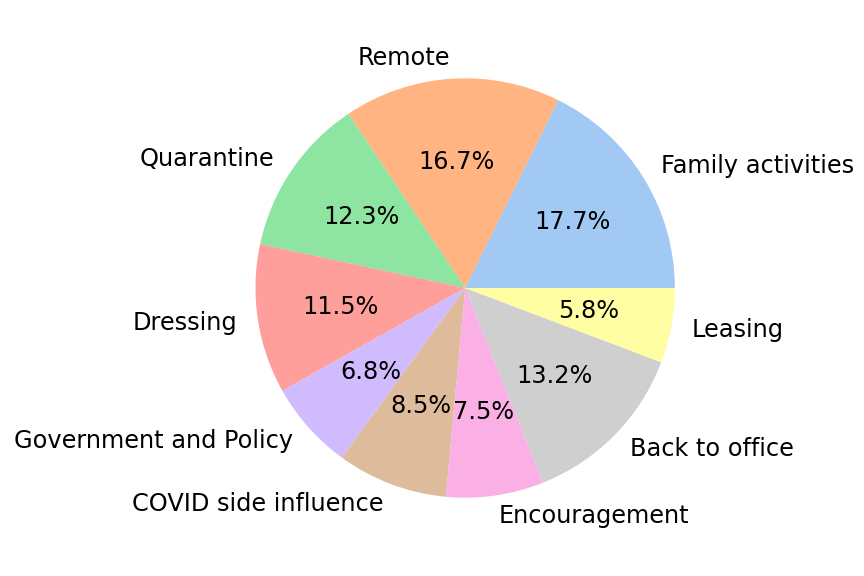}
    \caption{Topic distributions.}
    \label{fig:topic_dist}
\end{figure}

Figure~\ref{fig:topic_dist} shows the proportions of the topics. Topic 1 (Family activities) contains the keywords: “dog”, “wife” and “watch”, and constitutes 17.7\% of the total tweets. Topic 2 (Remote work/study) constitutes 16.7\%, where people mostly tweet about remote work and study. Topic 3 (Quarantine) contains the keywords “pandemic”, “force”, and “stay”. Topic 4 (Dressing) contains keywords “dress”, “wear”, and most of the tweets in this topic are talking about what they are wearing during WFH. Topic 5 (Government and policy) constitutes 6.8\%, which contains the keywords “money”, “job” and “force”. In this topic, many of the tweets mention governors and express their concerns about WFH-related policy. An example is: 

\begin{small}
{\tt \small @GovMurphy \#CancelRentNJ If NJ doesn't cancel rent, the consequences for everyone who can't work from home will be catastrophic. The bailout money will go straight to landlords instead of feeding people.}
\end{small}

Topic 6 (Covid side influence) contains keywords ``still'', ``job'', and constitutes 8.5\% of the total tweets. In this topic, people mostly complain about the influence of COVID-19, such as ``job changing'' and ``staying home for so long''. An example is:

\begin{small}
{\tt \small I agree. I think the pandemic has shown the disparity of digital access in rural areas, and libraries in these kinds of communities should take note of what these people need so they can provide better services when all of this is over. @Sonaite  @BakerChair \#SLIS752 \#752Diversity [QUOTE] @Sonaite @BakerChair @KaeliNWLib I think this pandemic has shown a spotlight on the digital divide that still exists. Schools are scrambling to provide "continuity of education" when children don't have access to devices and/or reliable Internet.}
\end{small}

Topic 7 (Encouragement) constitutes 7.5\% of the total tweets, where people express their support and inspire people to overcome the toughness together. One example is:

\begin{small}
{\tt \small If you are working out in this scary world today, my love to you. If you are working from home, my love to you. If you are out of work, my love to you. If you are lonely, my love to you. If you are sick, my love to you. If you are grieving, my love to you. My love. To you.}
\end{small}

Topic 8 (Back to office) contains keywords “back” and “office”, and constitutes 13.2\%. In this topic, people mostly tweet about their opinions towards the office, will “office” finally come back or when will be able to go back to office. Topic 9 (Leasing) contains keywords “year”, “lease”, “renew”, “office”, where people argue that some companies might not renew their office lease for the next year, because of how well WFH is working for these companies. 

\begin{figure}[htbp]
    \centering
    \includegraphics[width = 0.9\linewidth]{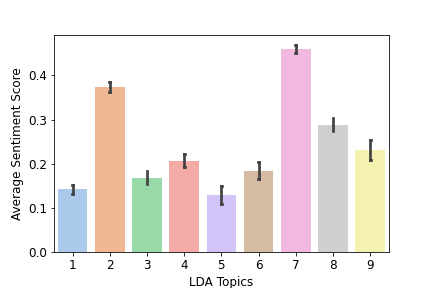}
    \caption{Average sentiment score of each topic.}
    \label{fig:topic_senti}
\end{figure}

Figure~\ref{fig:topic_senti} shows the average sentiment score of each topic. The average sentiment score of Topic 7 (Encouragement) is the highest (0.460), considered as the most positive topic, compared with Topic 5 (Government and policy), the least positive topic (0.129). As mentioned in the Feature Inference section, the average sentiment score of all the users in our dataset is 0.242. Among these 9 topics, all of them show a positive sentiment towards WFH. Meanwhile, Topic 2 (Remote work/study), Topic 7 (Encouragement) and Topic 8 (Back to office) are above the average; Topic 1 (Family activities), Topic 3 (Quarantine), Topic 4 (Dressing), Topic 5 (Government and Policy), Topic 6 (COVID side influence) and Topic 9 (Leasing) are below the average. 

\subsubsection{Money and jobs are discussed the most when the  government accounts are mentioned.} In Topic 5 (Government and policy), “money” and “job” are the most eye-catching keywords in LDA results. As we explore the tweets under this topic, we find there are a number of tweets mentioning government accounts and governor twitter accounts. An example is:

\begin{small}
{\tt \small @SenBobCasey @SenToomey @GovernorTomWolf supply chain workers discouraged. Work from home pieces of the chain are essential. TEMPORARY layoffs making more than I am to wait to get called back to work, where’s the incentive to work? Why aren’t we included in stimulus 2.0?}
\end{small}

In addition, we also find some tweets about money, where people are kind of worried about their financial status during the pandemic, such as losing money after lay-off by their companies. Based on these findings, since April is still at the early stage of WFH, economics would be a big deal for the government to handle during this period. 

\subsubsection{Family activities and Remote work/study conflicts in age groups.} 
\begin{figure*}[htbp]
    \centering
    \includegraphics[width =0.8 \linewidth]{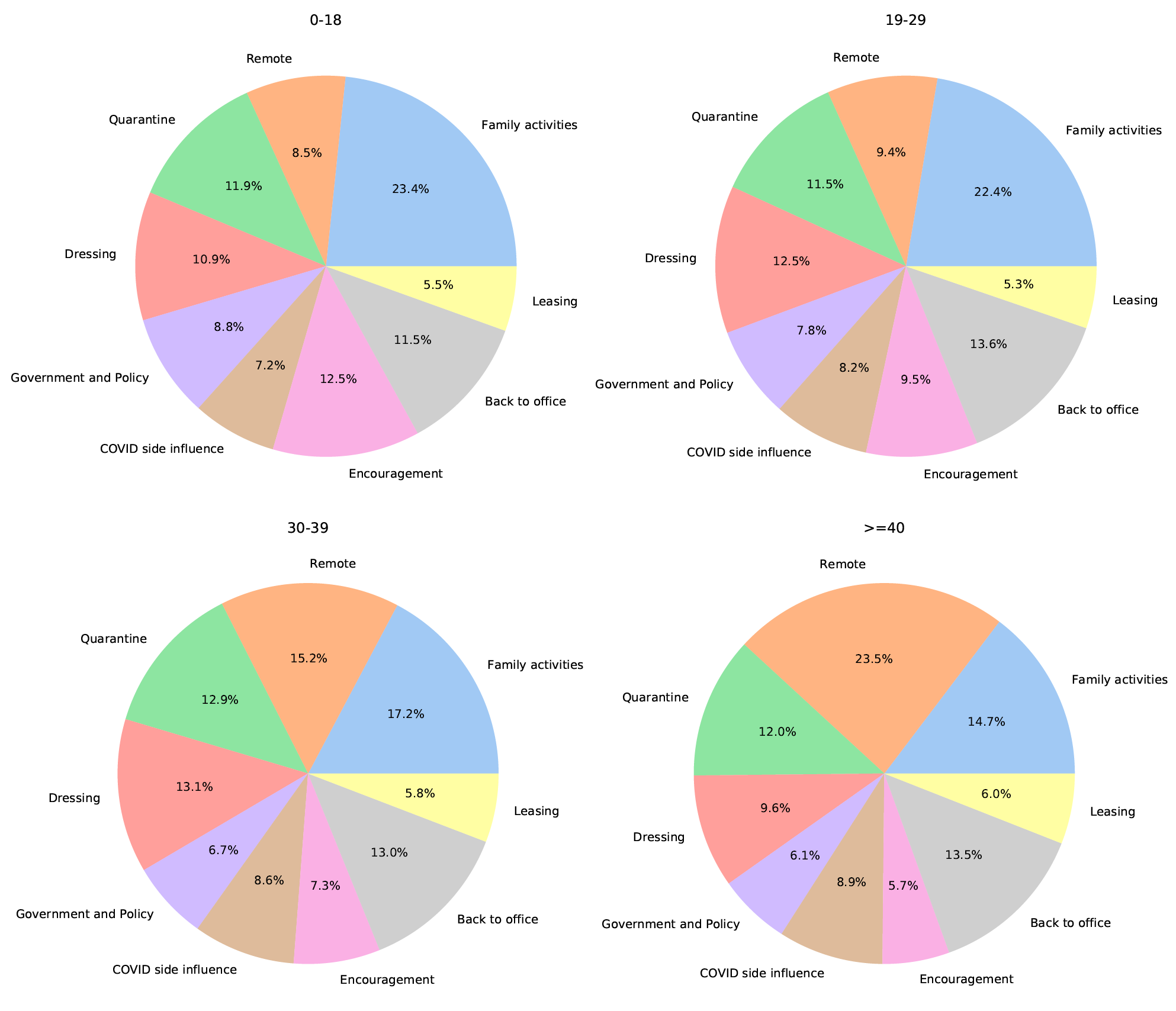}
    \caption{Topic distributions of age groups.}
    \label{fig:topic_dist_age}
\end{figure*}

According to Figure~\ref{fig:topic_dist_age}, as age goes up, the percentage of people tweeting about Remote work/study is rising; meanwhile, there are less people tweeting about family activities. For people aged from 0 to 18, there are 23.4\% of people tweeting about Topic 1 (Family activities) and 8.5\% about Topic 2 (Remote work/study). In the age group 19-29, there are 22.4\% of people talking about Family activities and 9.4\% about Remote work/study. For people in their 30s, there are 17.2\% on Family activities and 15.2\% on Remote work/study. For people older than 40, only 14.7\% of them tweet about Family activities, and 23.5\% on the topic of Remote work/study. In the aforementioned topic model summary, Topic 2 (Remote work/study) is a heavy work-related topic, which brings work-family conflicts on the table. \citet{frone1992prevalence} stated that family boundaries are more permeable than work boundaries. Interestingly, based on our findings, we can conclude that, in the environment of working from home, family boundaries are getting more permeable when it comes to older people. On average, >=40 age group accounts for 37.08\% of the study population. However, in Topic 1 (Family Activities), only 30.8\% of the tweets are from people older than 40; on the other hand, in Topic 2 (Remote work/study), 52.1\% of the tweets are coming from >=40 age group. These interesting patterns are consistent with our findings that family-work boundaries are getting weaker for older people.

\subsubsection{Superwomen in WFH.} In the previous section, we conclude that women show more positive attitudes than men to working from home. \citet{collins2020covid} found that might be because that women tend to reduce more work hours, and we also confirm this finding from the perspective of thematic analysis. By performing the goodness-of-fit test, we find that the topic distributions of men and women (Figure~\ref{tab:topic_dist_gender}) are significantly different (p < .001). According to the difference between the topic distributions of men and women, we think that fewer work hours allow women to spend more time accompanying children and taking care of family. Topic 1 (Family Activities) is one of the topics to which women pay more attention. As we dive deep into the tweets, we find that many tweets are about spending time with the children while working from home. An example is:

\begin{small}
{\tt \small That’d be me. I get to work from home and be with my kids. I’m loving every minute of this time with them!!}
\end{small}

\begin{figure*}[t]
    \centering
    \includegraphics[width = 0.8\linewidth]{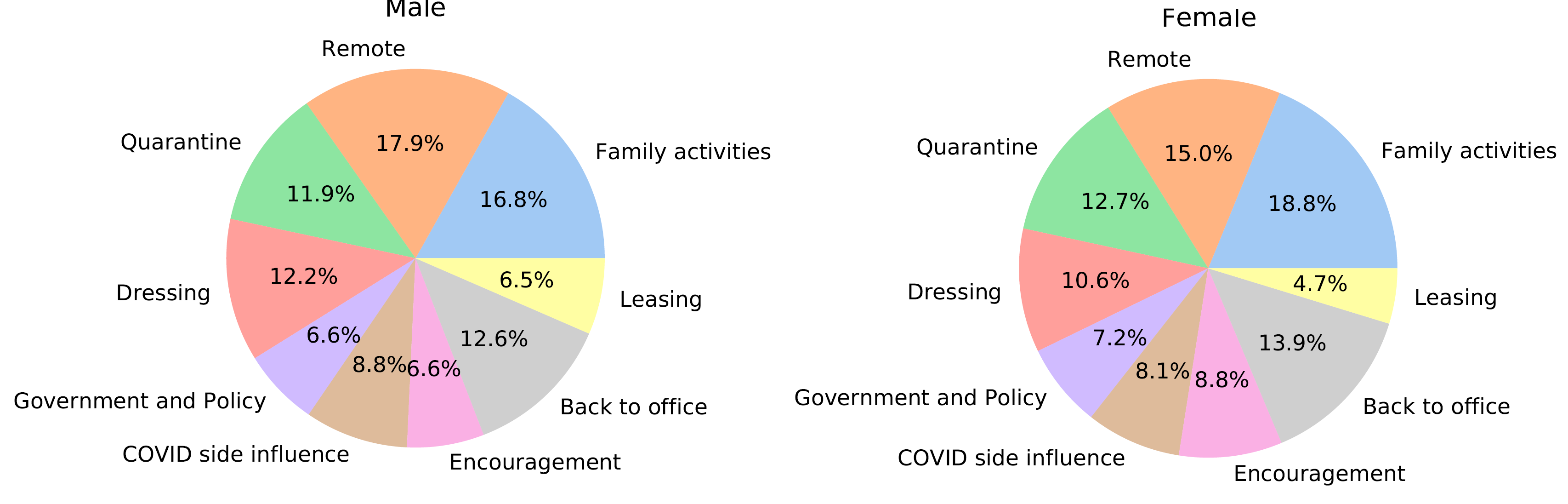}
    \caption{Topic distributions of gender groups.}
    \label{fig:topic_dist_gender}
\end{figure*}

\section{Discussion}
\subsection{Principal Results}
This study represents a large-scale quantitative analysis of public opinions on working from home during the COVID-19 pandemic. Through the
lens of social media, we find that gender and age are the most influential features to public opinions about WFH. After performing the ordinary least squares regression, we find that sentiment of working from home varies across user characteristics. In particular, women are more positive about working from home, which could be related to the change of working styles\footnote{https://www.fastcompany.com/90477102/4-ways-remote-work-is-better-for-women [Accessed March 24, 2021]} and fewer work hours compared to men~\cite{collins2020covid}. People in their 40s and older tend to be the most pro-WFH than other age groups. It could be due to the fact that people of those ages are the most vulnerable to COVID-19 while also the most difficult ones to get re-employed once they lost their jobs. They also need to work to mitigate the shrinkage of retirement savings that were invested in the sluggish stock market. People from high-income areas are more likely to have positive opinions about working from home, which echoes the findings of \citet{barrero2020working}.

These nuanced differences are supported by a more fine-grained topic analysis. At a higher level, we find that all the topics are showing a positive sentiment about WFH. However, people express a more negative sentiment towards the family activity and government. Within the topic of family activity, we notice that women pay more attention to family than men, and we find the superwomen in WFH. When people talk about government and policy, money and jobs are two major concerns. Furthermore, according to our analysis in age groups, we notice that the family work boundary is another issue that varies from different ages. As age increases, more people would like to talk about work rather than family, which implies that family boundaries are getting more permeable than work boundaries.

\subsection{Implications}
Barrero, Bloom and Davis have found that working from home will even stick after the pandemic ends~\cite{barrero2020working}. It is critical to understand public opinions on working from home to help improve their experience and design a more suitable and flexible work policy. Our paper suggests that there are nuanced differences across user characteristics. Policy-makers of the government and companies could design a more customized work policy to not only increase the work productivity but also improve the work satisfaction of their employees. It is also important to address the disparities related to working from home, which have been reported among different racial and socioeconomic groups~\cite{chowkwanyun2020racial, chang2020mobility}.

\subsection{Limitations}
Our study is focused on the relationship between user characteristics and the sentiment about working from home. However, user occupation can be included in the future analyses. Since the ability to work from home varies among different kinds of jobs~\cite{dingel2020many}, one potential hypothesis could be that people of different occupations hold different opinions about working from home, thus occupations would have an impact on the sentiment. In addition, there are also some limitations of only using Ethnicolr API to infer ethnicity. The Ethnicolr API is model trained on vote registration data from Florida state. First, using data from a single state (albeit a representative state) may not be ideal, since the pattern of the names can be different between states. For another, the training dataset is imbalanced (\textit{non-Hispanic White}: 8,757,268; \textit{non-Hispanic Black}: 1,853,690; \textit{Hispanic}: 2,179,106; \textit{Asian}: 253,808). Although these numbers are consistent with the facts of population distribution in the U.S., in the case of training an inference model, we still think using a more balanced dataset could provide a better result. 

\subsection{Conclusions}
This paper presents a large-scale social media-based study on who are more likely to tweet about working from home. By performing the ordinary least squares regression, we show how the sentiment of working from home varies across user characteristics. After conducting a content-based analysis, we dissect what Twitter users mainly talk about and how the content correlates with the sentiment of working from home. This paper contributes to a better understanding of public opinions on working from home during the COVID-19 pandemic and lends itself to making policies both at national and institution/company levels to improve the overall population's experience of working from home.

\subsection{Acknowledgements}
\subsubsection{Authors' Contributions.}
All authors conceived and designed the study. HL performed data collection. ZX and PL performed feature inference. PL conducted sentiment analysis. ZX applied LDA models. ZX and PL analyzed the data and wrote the majority of the manuscript. All authors edited the manuscript.

\subsection{Conflicts of Interest}
The authors declare that they have no competing interests.

\bibliography{main}

\begin{thebibliography}{26}
\providecommand{\natexlab}[1]{#1}
\providecommand{\url}[1]{\texttt{#1}}
\providecommand{\urlprefix}{URL }
\expandafter\ifx\csname urlstyle\endcsname\relax
  \providecommand{\doi}[1]{doi:\discretionary{}{}{}#1}\else
  \providecommand{\doi}{doi:\discretionary{}{}{}\begingroup
  \urlstyle{rm}\Url}\fi

\bibitem[{Barrero, Bloom, and Davis(2020)}]{barrero2020working}
Barrero, J.~M.; Bloom, N.; and Davis, S.~J. 2020.
\newblock Why Working From Home Will Stick.
\newblock \emph{University of Chicago, Becker Friedman Institute for Economics
  Working Paper} (2020-174).

\bibitem[{Bick, Blandin, and Mertens(2020)}]{bick2020work}
Bick, A.; Blandin, A.; and Mertens, K. 2020.
\newblock Work from home after the COVID-19 Outbreak .

\bibitem[{Blei, Ng, and Jordan(2003)}]{blei2003latent}
Blei, D.~M.; Ng, A.~Y.; and Jordan, M.~I. 2003.
\newblock Latent dirichlet allocation.
\newblock \emph{Journal of machine Learning research} 3(Jan): 993--1022.

\bibitem[{Boon-Itt and Skunkan(2020)}]{boon2020public}
Boon-Itt, S.; and Skunkan, Y. 2020.
\newblock Public perception of the COVID-19 pandemic on Twitter: sentiment
  analysis and topic modeling study.
\newblock \emph{JMIR Public Health and Surveillance} 6(4): e21978.

\bibitem[{Burger et~al.(2011)Burger, Henderson, Kim, and
  Zarrella}]{burger2011discriminating}
Burger, J.~D.; Henderson, J.; Kim, G.; and Zarrella, G. 2011.
\newblock Discriminating gender on Twitter.
\newblock In \emph{Proceedings of the 2011 Conference on Empirical Methods in
  Natural Language Processing}, 1301--1309.

\bibitem[{Carroll, Mostafa, and Thorne(2020)}]{carrollcovid}
Carroll, F.; Mostafa, M.; and Thorne, S. 2020.
\newblock COVID 19: WORKING FROM HOME: TWITTER REVEALS WHY WE’RE EMBRACING IT
  .

\bibitem[{Chang et~al.(2020)Chang, Pierson, Koh, Gerardin, Redbird, Grusky, and
  Leskovec}]{chang2020mobility}
Chang, S.; Pierson, E.; Koh, P.~W.; Gerardin, J.; Redbird, B.; Grusky, D.; and
  Leskovec, J. 2020.
\newblock Mobility network models of COVID-19 explain inequities and inform
  reopening.
\newblock \emph{Nature} 1--6.

\bibitem[{Chowkwanyun and Reed~Jr(2020)}]{chowkwanyun2020racial}
Chowkwanyun, M.; and Reed~Jr, A.~L. 2020.
\newblock Racial health disparities and Covid-19—caution and context.
\newblock \emph{New England Journal of Medicine} .

\bibitem[{Chung et~al.(2020)Chung, Seo, Forbes, and Birkett}]{chung2020working}
Chung, H.; Seo, H.; Forbes, S.; and Birkett, H. 2020.
\newblock Working from home during the COVID-19 lockdown: Changing preferences
  and the future of work .

\bibitem[{Collins et~al.(2020)Collins, Landivar, Ruppanner, and
  Scarborough}]{collins2020covid}
Collins, C.; Landivar, L.~C.; Ruppanner, L.; and Scarborough, W.~J. 2020.
\newblock COVID-19 and the gender gap in work hours.
\newblock \emph{Gender, Work \& Organization} .

\bibitem[{Dingel and Neiman(2020)}]{dingel2020many}
Dingel, J.~I.; and Neiman, B. 2020.
\newblock How many jobs can be done at home?
\newblock Technical report, National Bureau of Economic Research.

\bibitem[{Duong et~al.(2020)Duong, Luo, Pham, Yang, and Wang}]{duong2020ivory}
Duong, V.; Luo, J.; Pham, P.; Yang, T.; and Wang, Y. 2020.
\newblock The ivory tower lost: How college students respond differently than
  the general public to the covid-19 pandemic.
\newblock In \emph{2020 IEEE/ACM International Conference on Advances in Social
  Networks Analysis and Mining (ASONAM)}, 126--130. IEEE.

\bibitem[{Feng and Savani(2020)}]{feng2020covid}
Feng, Z.; and Savani, K. 2020.
\newblock Covid-19 created a gender gap in perceived work productivity and job
  satisfaction: implications for dual-career parents working from home.
\newblock \emph{Gender in Management: An International Journal} .

\bibitem[{Frone, Russell, and Cooper(1992)}]{frone1992prevalence}
Frone, M.~R.; Russell, M.; and Cooper, M.~L. 1992.
\newblock Prevalence of work-family conflict: Are work and family boundaries
  asymmetrically permeable?
\newblock \emph{Journal of organizational behavior} 13(7): 723--729.

\bibitem[{Hutto and Gilbert(2014)}]{hutto2014vader}
Hutto, C.; and Gilbert, E. 2014.
\newblock Vader: A parsimonious rule-based model for sentiment analysis of
  social media text.
\newblock In \emph{Proceedings of the International AAAI Conference on Web and
  Social Media}, volume~8.

\bibitem[{Jin et~al.(2010)Jin, Gallagher, Cao, Han, and Luo}]{jin2010wisdom}
Jin, X.; Gallagher, A.; Cao, L.; Han, J.; and Luo, J. 2010.
\newblock The wisdom of social multimedia: using flickr for prediction and
  forecast.
\newblock In \emph{Proceedings of the 18th ACM international conference on
  Multimedia}, 1235--1244.

\bibitem[{Lyu et~al.(2020)Lyu, Wang, Wu, Duong, Zhang, Dye, and
  Luo}]{Lyu2020.12.12.20248070}
Lyu, H.; Wang, J.; Wu, W.; Duong, V.; Zhang, X.; Dye, T.~D.; and Luo, J. 2020.
\newblock Social Media Study of Public Opinions on Potential COVID-19 Vaccines:
  Informing Dissent, Disparities, and Dissemination.
\newblock \emph{medRxiv} .

\bibitem[{Palumbo(2020)}]{palumbo2020let}
Palumbo, R. 2020.
\newblock Let me go to the office! An investigation into the side effects of
  working from home on work-life balance.
\newblock \emph{International Journal of Public Sector Management} .

\bibitem[{Rai{\v{s}}ien{\.e} et~al.(2020)Rai{\v{s}}ien{\.e}, Rapuano,
  Varkulevi{\v{c}}i{\=u}t{\.e}, and Stachov{\'a}}]{raivsiene2020working}
Rai{\v{s}}ien{\.e}, A.~G.; Rapuano, V.; Varkulevi{\v{c}}i{\=u}t{\.e}, K.; and
  Stachov{\'a}, K. 2020.
\newblock Working from Home—Who is Happy? A Survey of Lithuania’s employees
  during the COVID-19 quarantine period.
\newblock \emph{Sustainability} 12(13): 5332.

\bibitem[{Sood and Laohaprapanon(2018)}]{sood2018predicting}
Sood, G.; and Laohaprapanon, S. 2018.
\newblock Predicting race and ethnicity from the sequence of characters in a
  name.
\newblock \emph{arXiv preprint arXiv:1805.02109} .

\bibitem[{Talbot, Charron, and Konkle(2021)}]{talbot2021feeling}
Talbot, J.; Charron, V.; and Konkle, A. 2021.
\newblock Feeling the Void: Lack of Support for Isolation and Sleep
  Difficulties in Pregnant Women during the COVID-19 Pandemic Revealed by
  Twitter Data Analysis.
\newblock \emph{International Journal of Environmental Research and Public
  Health} 18(2): 393.

\bibitem[{{U.S. Census Bureau}(2019)}]{census2019}
{U.S. Census Bureau}. 2019.
\newblock Qucik Facts Table.

\bibitem[{Wang et~al.(2019)Wang, Hale, Adelani, Grabowicz, Hartman, Fl{\"o}ck,
  and Jurgens}]{wang2019demographic}
Wang, Z.; Hale, S.; Adelani, D.~I.; Grabowicz, P.; Hartman, T.; Fl{\"o}ck, F.;
  and Jurgens, D. 2019.
\newblock Demographic inference and representative population estimates from
  multilingual social media data.
\newblock In \emph{The World Wide Web Conference}, 2056--2067. ACM.

\bibitem[{Wu, Lyu, and Luo((in-press))}]{wu2021characterizing}
Wu, W.; Lyu, H.; and Luo, J. (in-press).
\newblock Characterizing Discourse about COVID-19 Vaccines: A Reddit Version of
  the Pandemic Story.
\newblock \emph{Health Data Science} .

\bibitem[{Yeung, Lai, and Luo(2020)}]{yeung2020face}
Yeung, N.; Lai, J.; and Luo, J. 2020.
\newblock Face off: Polarized public opinions on personal face mask usage
  during the COVID-19 pandemic.
\newblock In \emph{2020 IEEE International Conference on Big Data (Big Data)},
  4802--4810. IEEE.

\bibitem[{Zhang et~al.((in-press))Zhang, Lyu, Liu, Zhang, Wang, and
  Luo}]{zhang2021monitoring}
Zhang, Y.; Lyu, H.; Liu, Y.; Zhang, X.; Wang, Y.; and Luo, J. (in-press).
\newblock Monitoring Depression Trends on Twitter During the COVID-19 Pandemic:
  Observational Study.
\newblock \emph{JMIR Infodemiology} .

\end{thebibliography}

\subsection{Abbreviations}
ACS: American community survey\\
API: application programming interface\\
COVID-19: coronavirus disease 2019\\
HR: human resources\\
LDA: Latent Dirichlet Allocation\\
NLP: natural language processing\\
NLTK: natural language toolkit\\
RQ: research question\\
VARDER: valence aware dictionary and sentiment reasoner\\
WFH: work from home

\end{document}